\documentclass[namedreferences]{solarphysics}
\usepackage[hyperref,optionalrh]{spr-sola-addons} 
\usepackage{graphicx} 
\usepackage{color}   
\usepackage{breakurl}
\ifx \doiurl    \undefined \def \doiurl#1{\href{http://dx.doi.org/#1}{\textsf{DOI}}}\fi
\ifx \arxivurl  \undefined \def \arxivurl#1{\href{http://arxiv.org/abs/#1}{\textsf{arXiv: #1}}}\fi
\ifx\urlurl    \undefined \def\urlurl#1{\href{http://#1}{\textsf{#1}}}\fi
\begin{document}

\begin{opening}

\title{Large-Scale Vortex Motion and Multiple Plasmoid Ejection Due to Twisting Prominence Threads and Associated Reconnection}
\author[addressref=aff1,corref,email={sudheerkm.rs.phy16@itbhu.ac.in}]{\inits{S.K.}\fnm{Sudheer K.}~\lnm{Mishra}\orcid{https://orcid.org/0000-0003-2129-5728}}
\author[addressref=aff1]{\inits{A.}\fnm{A.K.}~\lnm{Srivastava}}
\author[addressref={aff2,aff3}]{\inits{P.F.}\fnm{P.F.}~\lnm{Chen}}
\address[id=aff1]{Department of Physics, Indian Institute of Technology (BHU), Varanasi-221005, India.}
\address[id=aff2]{School of Astronomy \& Space Science, Nanjing University, Nanjing 210023, China.}
\address[id=aff3]{Key Laboratary of Modern Astronomy and Astrophysics (Nanjing University), Ministry of Education, China.}
\runningauthor{S.K. Mishra et al.}
\runningtitle{Plasmoid Enhanced Reconnection in Solar Prominence}
\begin{abstract}
We analyze the characteristics of a quiescent polar prominence using the \textit{Atmospheric Imaging Assembly} (AIA) onboard the \textit{Solar Dynamics Observatory} (SDO). Initially, small-scale barb-like structures are evident on the solar disk, which firstly grow vertically and thereafter move towards the south-west limb. Later, a spine connects these barbs and we observe apparent rotating motions in the upper part of the prominence. These apparent rotating motions might play an important role for the evolution and growth of the filament by transferring cool plasma and magnetic twist. The large-scale vortex motion is evident in the upper part of the prominence, and consists of a swirl-like structure within it. The slow motion of the footpoint twists the legs of the prominence due to magnetic shear, causing two different kinds of magnetic reconnection. The internal reconnection is initiated by a resistive tearing-mode instability, which leads to the formation of multiple plasmoids in the elongated current sheet. The estimated growth rate was found to be 0.02\,--\,0.05. The magnetic reconnection heats the current sheet for a small duration. However, most of the energy release due to magnetic reconnection is absorbed by the surrounding cool and dense plasma and used to accelerate the plasmoid ejection. The multiple plasmoid ejections destroy the current sheet. Therefore, the magnetic arcades collapse near the X-point. Oppositely directed magnetic arcades may reconnect with the southern segment of the prominence and an elongated thin current sheet is formed. This external reconnection drives prominence eruption.\\

{\textit{Keywords}: Prominence, Vortex-motion, Reconnection, Corona}
\end{abstract}
\end{opening}
\section{Introduction}
                          The solar corona consists of large-scale sheared or twisted magnetic fields that lie above the polarity inversion line. These sheared magnetic fields are sometimes associated with solar filaments (or so-called prominences). Solar prominences are comparatively cool and dense plasma structures suspended in the rarified and hotter million-degree corona with the help of magnetic field (e.g. Labrosse {\it et al.}, 2010; Mackay {\it et al.}, 2010; Parenti, 2014). On the basis of magnetic-field strength and the location of the solar prominences, they may be classified as active-region, quiescent, and intermediate prominence (e.g. Tandberg-Hanssen, 1998; Mackay {\it et al.}, 2010; Parenti, 2014). The prominence formation, stability, and lifetime have been governed by the magnetic field (e.g. Priest, 1989; Mackay {\it et al.}, 2010). Various observational and theoretical models have been used to understand the magnetic structures of the solar prominences. The most recognized models are ``sheared arcade model'' (e.g. Antiochos, Dahlburg and Klimchuk, 1994; DeVore and Antiochos, 2000; Aulanier and Schmieder, 2002) and ``flux rope model'' (e.g. Kuperus and Raadu, 1974; Priest, 1989; van Ballegooijen and Martens, 1989; Rust and Kumar, 1994; Low and Hundhausen, 1995; Aulanier {\it et al.} 1998; van Ballegooijen 2004; Gibson and Fan, 2006; Dudik {\it et al.}, 2008). It was found that 89\,\% of filaments are supported by flux ropes, with the rest by sheared arcades (Ouyang {\it et al.}, 2017). In the sheared-arcade model, only sheared magnetic-field lines are embedded in a much less sheared envelope field. In the flux-rope model, the core magnetic field is twisted and in a helical configuration. Some additional models have also been introduced to discuss the stability of hedgerow and polar prominences. The most known models are the ``tangled field model'' for hedgerow prominence (van Ballegooijen and Cranmer, 2010) and the ``flux linkage model'' (Martens and Zwaan, 2001) for the development of polar prominences. A hedgerow prominence consists of multiple vertical threads and they are sustained by tangled magnetic fields (van Ballegooijen and Cranmer, 2010). According to the flux-linkage model, two unconnected bipoles reconnect due to the flux convergence and reconnection near the polarity inversion line, which creates longer magnetic loops with dips. Similar processes occur at multiple times and then the prominence develops highly sheared longer structures (e.g. Martens and Zwaann, 2001; Panesar {\it et al.}, 2014).   \\
 \begin{figure*}
\hspace{-1.0cm}
\includegraphics[scale=1.0,angle=0,width=13.0cm,height=13.0cm,keepaspectratio]{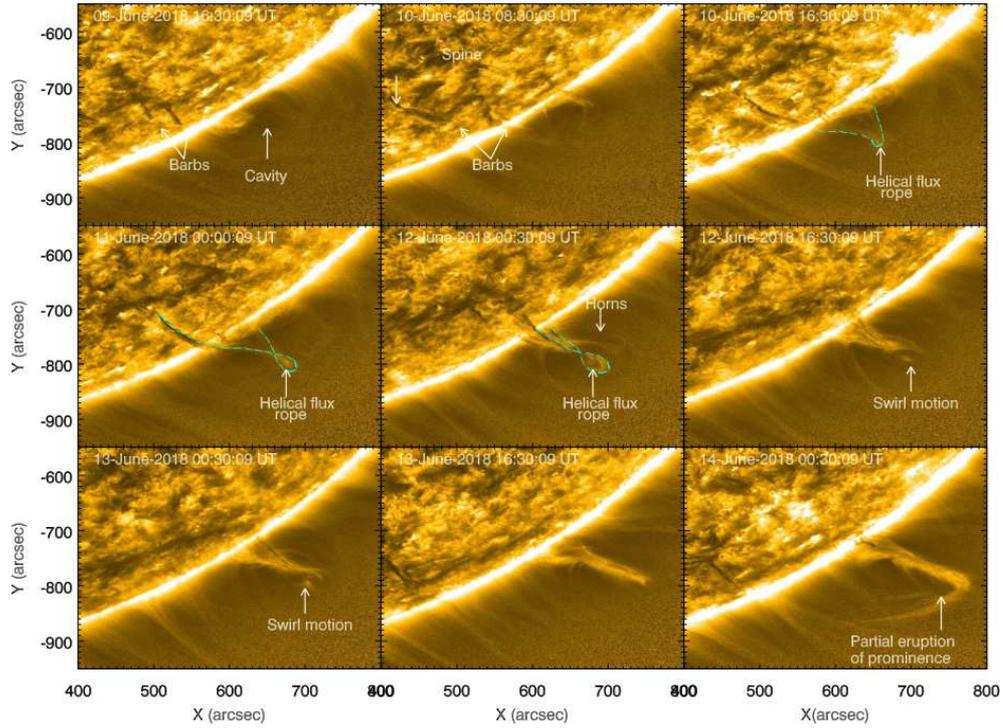}
\caption{The normalized multi-gaussian fitted SDO/AIA 171 {\AA} images showing the evolution of solar tornado-like motions related to the polar prominence during 9\,--\,14 June 2018. Two dark pillars indicate the prominence barbs and the horizonatl structure that connects these barbs in the prominence spine. Several vertical threads and U-shaped horn-like structures have also evident (middle panel). Rotating plasma structure indicates that the growth of the prominence is related to the tornado-like motions. The electronic supplementary materials (movie1.mp4) shows the evolution of a helical flux rope, apparent rotating plasma, swirling motion on the top of the prominence. It runs from 9 June 2018 at 08:30 UT to 14 June 2018 07:30 UT and includes the barbs, spine, swirl motion, and the related prominence. }
\end{figure*}
Solar filaments have two main structures, the spine and several barbs. In the EUV emissions images of polar-crown filaments we mainly see barbs, which are known as filament legs and are observed as dark pillars (e.g. Schmieder {\it et al.}, 2010; Dudik {\it et al.}, 2012; Li and Zhang, 2013). The evolution of multiple barbs in a polar prominence was observed by \textit{Solar Dynamics Observatory/Atmospheric Imaging Assembly} (SDO/AIA) (Li and Zhang, 2013). The kinematics of these barbs is highly influenced by the photospheric magnetic field. The tornado-like structure is useful to support the prominence plasma (Luna, Karpen, and DeVore, 2015). They found that the Lorentz force supports the barbs plasma if the magnetic field is sufficiently twisted. Solar tornadoes are magnetized plasma and their footpoints are rooted at the photoshpere. They were sometimes suggested to be rotating vertical magnetized structures driven by the vortex-like motions in the photosphere (Su {\it et al.}, 2012). These tornadoes act as the channels of energy and mass transfer into the corona (Wedemeyer-Bohm {\it et al.}, 2012). Recent high-resolution observations taken from SDO suggest that the tornado-like prominences are associated with the apparent rotational motion of the helical structures (Li {\it et al.}, 2012; Su {\it et al.}, 2014; Panesar {\it et al.}, 2013; Yang {\it et al.}; 2018). In contrast, some authors suggest that the tornado-like motion of the prominence is an illusion due to the line of sight projection effect. The apparent rotational motion can arise due to the counter-streaming flow or oscillations in the prominence spine and barbs in the plane of the sky (Panasenco, Martin and Velli, 2014). Anti-symmetric Doppler velocity pattern along the axis of prominence legs was found only in the coronal plasma with a temperature above 1\,MK (Su {\it et al.}, 2014). Doppler maps of Mg {\sc II} K, which is formed in the frontmost layers of prominence plasma, do not show anti-symmetric or split pattern and do not support any of the rotating ``tornado'' models (Levens {\it et al.}, 2016a). Instead, the horizontal counter-streaming of prominence threads connecting two prominence legs is observed in Mg {\sc II} K line (Levens {\it et al.}, 2016b). The plasma behavior associated with the internal magnetic field of a tornado-like prominence was investigated and it was found that the quasi-vertical magnetic flux that a tornado in a helical flux rope needs is not consistent with the totally horizontal magnetic field measured in prominence legs (Levens {\it et al.}, 2016a, 2016b). \\

Tornado-like motions play an important role in suppling the mass and magnetic shear into the prominence. The rotating motion of barbs may transfer the shear and cool plasma into a prominence (Su {\it et al.}, 2014). The sheared prominence legs may consist of oppositely directed magnetic fields, which reconnect over the neutral line. The oppositely directed magnetic field of the filament legs leads to two-stage magnetic reconnection, as we report observationally in the present work. The reversal of magnetic shear may lead to two-stage of magnetic reconnection (Kusano {\it et al.}, 2004). The first stage of magnetic reconnection (internal reconnection) is governed by the resistive-tearing-mode instability, which is dominated by multiple plasmoid ejections. Yohkoh data provided the first observational evidence of X-ray plasmoid ejections. Long before the impulsive phase, the plasmoids may be ejected slowly and during the impulsive phase, they may accelerate rapidly (e.g. Ohyama and Shibata, 1997, 1998; Tsuneta, 1997; Liu, Chen and Petrosian, 2015). The apparent rise velocity of the flares and reconnection rate is positively correlated (Shibata {\it et al.}, 1995). The ejected H$\alpha$ plasmoids/filaments and thermal energy density of X-ray loop arcades are positively correlated. The ratio between the inflow speed and the plasmoid velocity is equivalent to the reconnection rate. Using the above results, Shibata (1996, 1997) suggested a new model for the magnetic reconnection i.e. plasmoid-induced reconnection, which is an extension of the standard Carmichael--Sturrock--Hirayama--Kopp--Pneuman (CSHKP) solar flare model. Inside the current sheet, the ejected plasmoids act as a resistance for the magnetic reconnection. Therefore, these plasmoids act as storage of magnetic energy. These plasmoids induce a strong inflow when they ejected from the current sheet. The current sheet in a solar flare consists of hot X-ray plasmoid ejection and cool H$\alpha$ plasmoid ejection (Ohyama and Shibata, 2007). \\
A solar filament reconnects with the nearby coronal loops and forms an elongated current sheet (e.g. Li {\it et al.}, 2016; Xue {\it et al.}, 2016; Dai {\it et al.}, 2018). Magnetic reconnection can also happen in the lower solar atmosphere (photosphere and chromosphere), which is made of collision-dominant, partially ionized, dense, and cool plasma. The energy released by reconnection is found to be absorbed by the cold plasma (e.g. Litvinenko, 1999; Chen, Fang and Ding, 2001; Chen and Ding, 2006; Litvinenko, Chae, and Park, 2007; Xue {\it et al.}, 2018). Solar prominences have similar types of physical properties (partially ionized, collision dominant, dense and cool plasma). Therefore, most of the energy release during the forced magnetic reonnection can be absorbed by the prominence as well (Srivastava {\it et al.}, 2019). Murphy {\it et al.} (2012) investigated the asymmetric magnetic reconnection between solar flares and CME. The asymmetric inflows induce the ejeted plasmoids to roll along the flux-rope axis. In the elongated current sheet, fractional and turbulent magnetic reconnection was revealed (e.g. Shibata and Takasao, 2016; Cheng {\it et al.}, 2018). Kusano et al. (2003) investigate the sheared magnetic arcade and suggest that the magnetic reconnection occurs at the shear inversion line. Footpoints of the magnetic arcades sheared in opposite directions above the PIL. Magnetic shear above polarity inversion line (PIL) may be responsible for two different kinds of magnetic reconnection (Kusano {\it et al.}, 2004). The second stage of the magnetic reconnection (external reconnection) triggers the prominence eruption in the above mentioned case.\\
\begin{figure*}
\includegraphics[scale=1.0,angle=0,width=12.0cm,height=12.0cm,keepaspectratio]{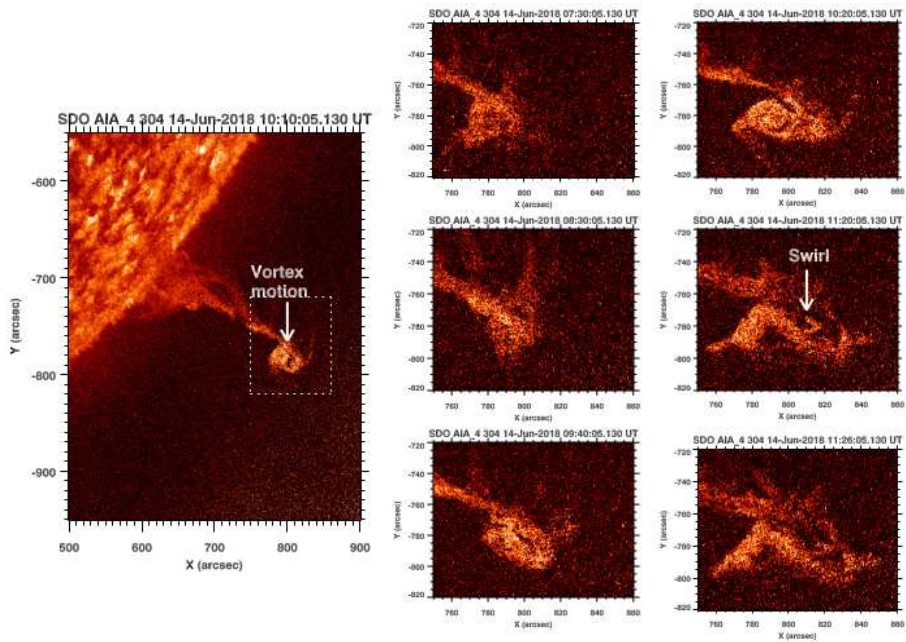}
\caption{Left panel: Full field of view (FOV) of a polar prominence observed by SDO/AIA 304 {\AA} on 14 June 2018. The sub-region (white box) consists of a large-scale vortex motion and a swirl-like structure. Right panel: The spatio-temporal evolution of the large-scale vortex motion. The cool plasma and photospheric vortices (large-scale vortex) channeled through the prominence body. The animation (movie2.mp4) of the SDO/AIA 304 {\AA} image shows the full evolution of large-scale vortex motion and swirl-like structure within it. It runs from 09:20 UT to 11:50 UT.}
\end{figure*}
In this article, we analyze the observational data to investigate the detailed process associated with magnetic reconnection between a prominence and the ambient corona. The present work is useful to understand the dynamics of the polar prominence, the association of tornado-like motion with the prominence growth and onset of the large-scale vortex motion. Observational data and analysis are presented in Sect.~2. Sect.~3 describes the observational results corresponding to development of the prominence, plasmoid enhanced reconnection, and the eruption of the prominence. In Sect.~4, discussion and conclusions are presented.\\
\begin{figure*}
\includegraphics[scale=1.0,angle=0,width=12.0cm,height=12.0cm,keepaspectratio]{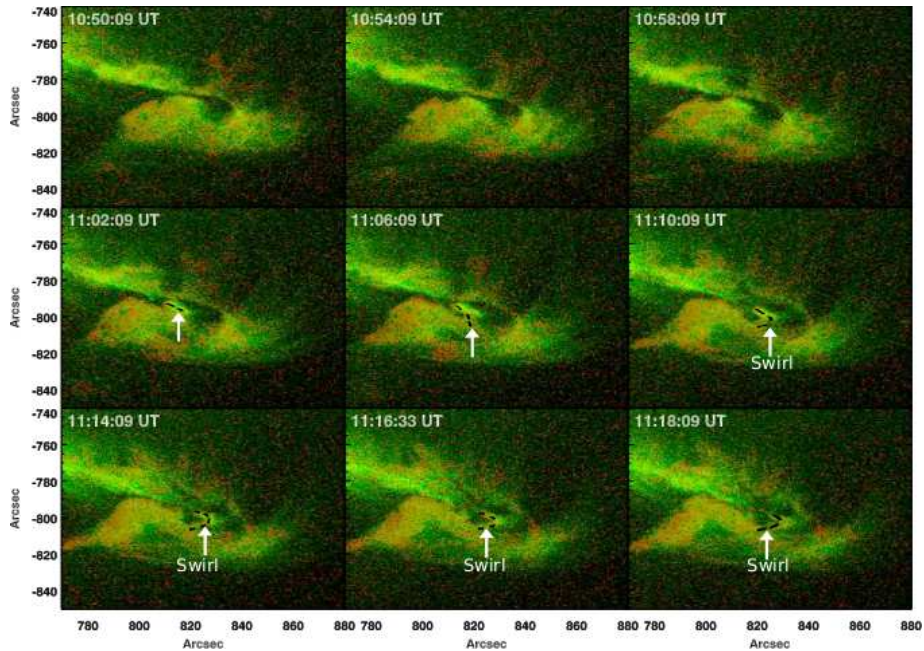}
\caption{The region of interest (ROI; white box in Figure~2) is displayed in the SDO/AIA (171+304) {\AA} composite images. The composite images display the large-scale vortex motion and swirl-like structure in the upper part of the prominence. We observed the spatio-temporal evolution of swirl indicated by white arrow within the prominence.}
\end{figure*}

\section{Observational Data and Analysis}
We use high-resolution data of the {\textit Atmospheric Imaging Assembly} (AIA: Lemen {\it et al.}, 2012) onboard the {\textit Solar Dynamics Observatory} (SDO: Pesnell, Thompson, and Chamberlin, 2012). The AIA telescope provides full-disk images of the solar atmosphere (photosphere, chromosphere and corona) in seven extreme ultraviolet channels (94 {\AA}, 131 {\AA}, 171 {\AA}, 193 {\AA}, 211 {\AA}, 304 {\AA}, 335 {\AA}), two ultraviolet channels (1600 {\AA}, 1700 {\AA}), and one visible-imager channel (4500 \AA). The AIA data have a 1.5 arcsec spatial resolution and 12 second temporal resolution for the EUV channels. We use the {AIA 94 {\AA}, 131 {\AA}, 171 {\AA}, 193 {\AA}, 211 {\AA}, 304 {\AA}, and 335 {\AA} images for our observations. The basic calibration and analysis are performed by using the IDL routines aia\_prep.pro and aia\_offlimb.pro, which are available in the Solarsoft library. \\
\begin{figure*}
\includegraphics[scale=1.0,angle=0,width=12.0cm,height=12.0cm,keepaspectratio]{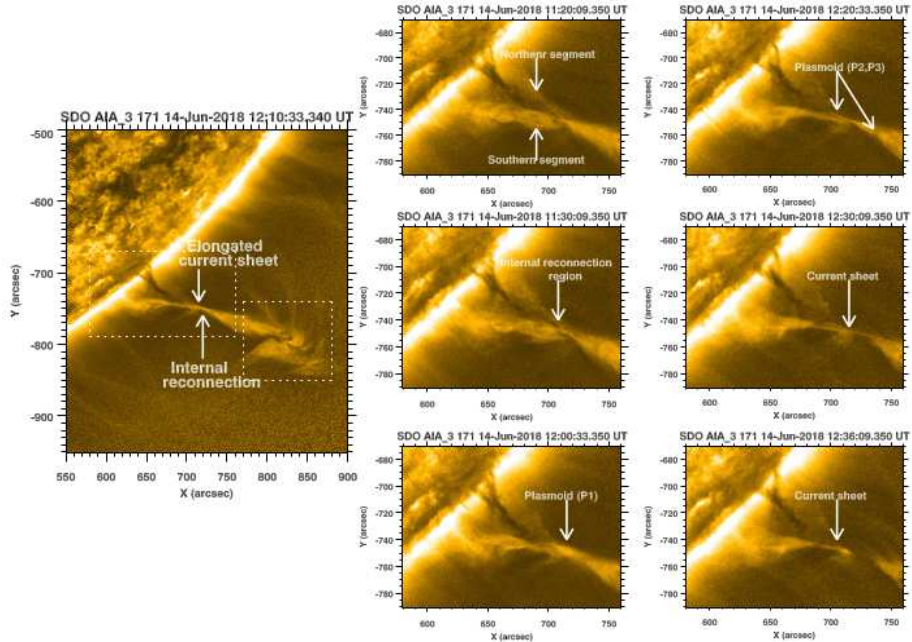}
\caption{Left panel: The region of interest (ROI) of the polar prominence legs observed by SDO/AIA 171 {\AA} on 14 June 2018. The oppositely directed prominence legs reconnect above the PIL. Right panel: Resistive-tearing-mode instability initiate the formation of multiple plasmoids within the current sheet. We observed three ejecting plasmoids and a bright cusp above the prominence legs (indicated by white arrows). An animation (movie3.mp4) show that oppositely directed prominence legs reconnect over a polarity inversion line (PIL) and initiate the plasmoid enhanced reconnection. We assume that the PIL is lying between the prominence legs. The animation runs from 11:00 UT to 13:00 UT.}
\end{figure*}
The thermal properties of the elongated current sheet and plasmoids embedded in the dynamical prominence system are understood by estimating the differential emission measure (DEM). We map the DEM in the temperature range between log $T$=5.6\,--\,6.7 using six AIA filters, i.e 94 {\AA}, 131 {\AA}, 171 {\AA}, 193 {\AA}, 211 {\AA}, 335 {\AA}. The method developed by Cheung {\it et al.}, (2015) is adopted to calculate the DEM from the elongated current sheet and ejected plasmoids within the prominence system. This method adopts the concept of the sparse inversion, which uses the ``simplex'' function to minimize the total emission. The positive solution has been found by adopting the sparse inversion method, which lies between max(0, I-tol) and (I+tol), where tol is the tolerance into the reconstructed intensities. For the sparse inversion, we divide the range of temperature between log $T$(K)=5.0\,--\,7.5 with 25 temperature bins at log $T$(K)=0.1 intervals. The AIA images are processed using the normalized multiscale Gaussian filter which is used to enhance the fine structures of eruptive prominences and magnetic arcades (Morgan and Druckmuller, 2014). \\
Using the above observational data and different observational techniques, we analyze a polar prominence, which is associated with tornado-like motions in the solar corona. The small-scale dark pillars (barbs) interact with each other and form a large-scale prominence. This prominence shows various internal dynamics for more than four days at different spatio-temporal scales. We use nine hours of AIA emission data starting on 14 June 2018 at 09:20 UT. During the evolution of the prominence, we observed a large-scale vortex motion, plasmoid enhanced reconnection, and associated prominence eruption.\\
\begin{figure*}
\hspace{-1cm}
\includegraphics[scale=1.0,angle=0,width=13.0cm,height=13.0cm,keepaspectratio]{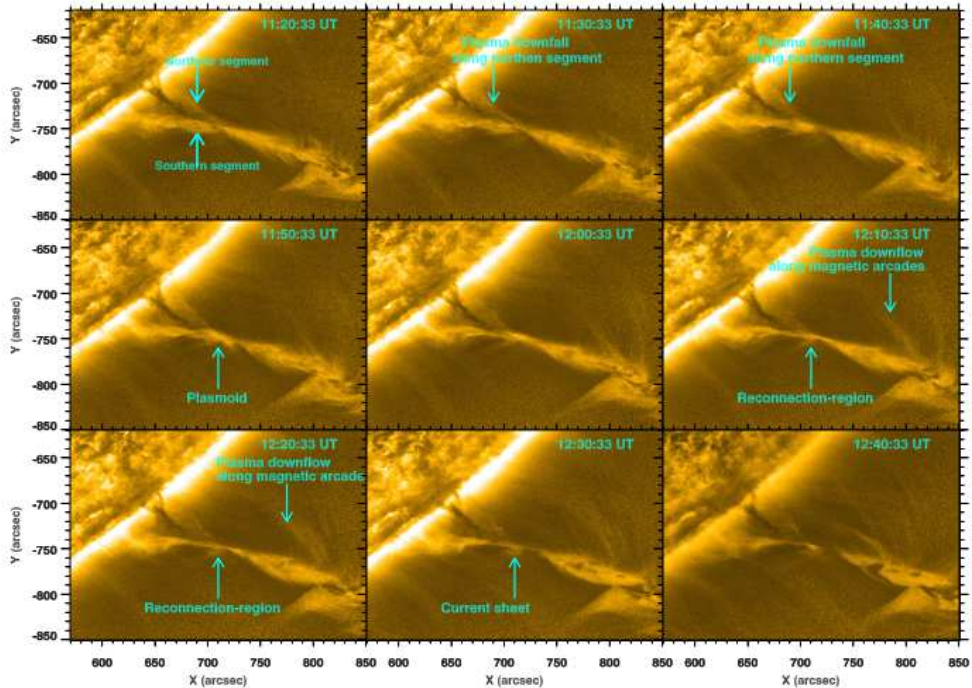}
\caption{The normalized multi-gaussian fitted SDO/AIA 171 {\AA} images display the FOV of polar prominence. Upper panel: Southern segment of the prominence is rising while northern segment is falling along prominence legs. The oppositely directed prominence legs reconnect above the PIL. Multple plasmoid ejection is also observed during the reconnection. These plasmoid passes through the main body of the prominence. Simultaneously, plasma downfall was also evident along the magnetic arcades.}
\end{figure*}
\section{Observational Results}
As mentioned above, we observed a polar crown prominence that erupted on 14 June 2018 at 13:35 UT, as indicated by different AIA wavebands. Before the eruption of this prominence, several internal dynamics are observed. The polar prominence initially appeared on 9 June 2018. We use high-cadence and high-resolution AIA 171 {\AA} emission images to investigate the internal dynamics of prominence and its associated thread-like structures. The subsection below outlines the step-wise development of the eruptive prominence.\\ 

\subsection{Evolution of the Prominence}
 Before the filament eruption, several types of internal dynamics (prominence barbs, spine, vertical threads, swirl motion, horn-like structure, plasmoid-enhanced reconnection, and associated inflows and outflows) are observed. The prominence consists of three structural components, i.e. barbs, spine, and prominence legs. The barbs and the spines are vertical and horizontal parts of the prominence. Plasma flows and the development of the prominence are governed by barbs, which may be rooted in the network boundaries (e.g. Płocieniak and Rompolt, 1973; Lin {\it et al.}, 2005). Vortex motion is also evident at the network boundaries. We observe here the step-wise evolution of the tornado-like prominence and its dynamics (Figure~1). However, in this article, we are not interested in understanding the development of the tornado and its internal dynamics. Instead, we are interested in the later phase of this prominence when it is in an eruptive stage. Two barbs appear on 9 June 2014 (Figure~1, top panel). These structures grow horizontally and vertically and one day later on 10 June 2018, a spine connects these barbs. In the later stage of the development, we observe that vertical threads appeare in the upper part of the prominence (Figure~1, movie1.mp4, middle panel). These vertical threads are associated with the quiet-region of the Sun. Horn-like structures are also observed on the top of these vertical threads (Figure~1, lower middle panel). The horns connect cooler and denser prominence plasma with the rarefied and hotter corona. The morphology of the horns indicates that they are field aligned structures, and therefore they may be used to understand the magnetic configuration of prominence-corona transition region (e.g. Berger {\it et al.}, 2012; Liu, Berger and Low, 2012; Luna {\it et al.}, 2012; Schmit and Gibson, 2013; Schmit {\it et al.}, 2013; Wang {\it et al.}, 2016). A cavity lies above the prominence vertical thread (Figure~1, middle panel). Initially, thin helical structure starts to rise in the cavity (Figure~1, top panel). Some plasma material falls along the horns and the helical field lines. The falling plasma material follows the path of the helical field lines, which causes the swirling motion on the top of the prominence (Figure~1, middle panel). Cool prominence associated plasma shows shearing or rotational motions on the top of the vertical threads. The plasma flow occurs via the horns, which acts as a channel to transfer mass and shearing to the prominence. Several types of internal dynamics and the partial eruptions are observed during the growth of this prominence (Figure~1, bottom panel, movie1.mp4). We do not aim to understand these dynamics and partial eruptions. On the contrary, we focus on the dynamics of the prominence on 14 June 2018, before and after its eruption, and the associated plasma outflows.\\
\subsection{Large-Scale Vortex Motion}
The eruptive prominence is associated with twisting/swirling motions. These twisting/swirling motions may act as a channel to transport the cool and dense plasma and twist into the prominence spine. On 14 June 2018 at 09:30 UT, sheared prominence legs transport the large-scale vortices (swirls) to the upper part of the prominence.
\begin{figure*}
\hspace{-1.0cm}
\includegraphics[scale=1.0,angle=0,width=13.0cm,height=13.0cm,keepaspectratio]{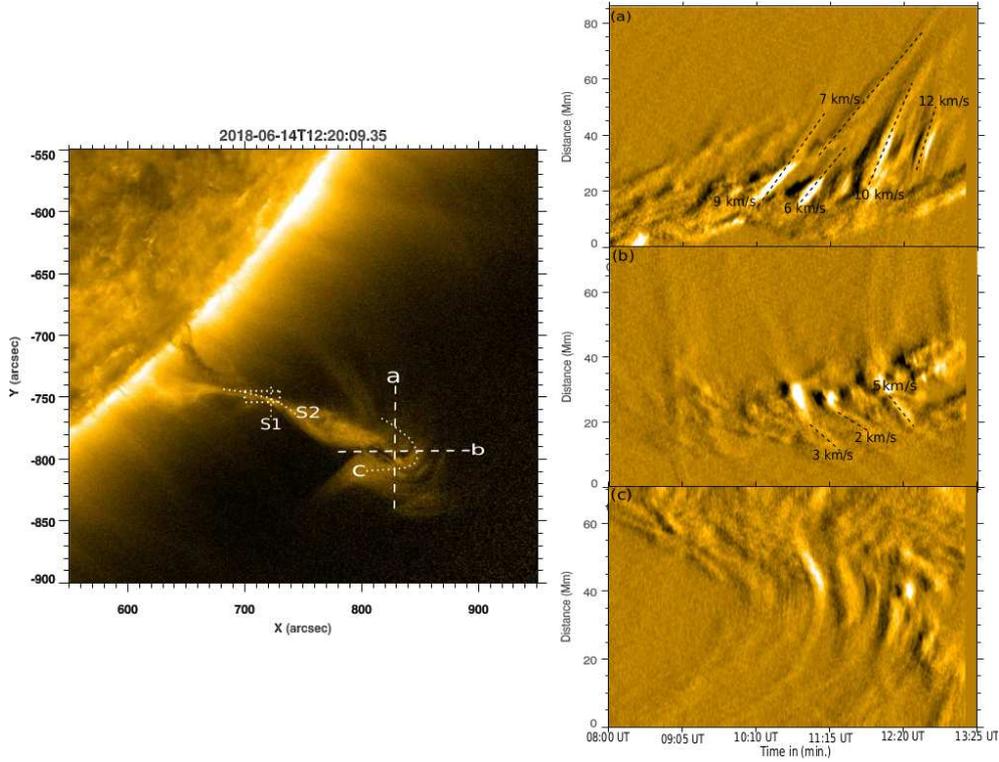}
\caption{The full FOV of the polar prominence in AIA 171 {\AA} emission imager to analyses the inflow, outflow, and nature of EUV intensity pattern at the reconnection site. We have estimated the EUV emission from the selected white-box region. Two paths (slit S1 and S2) has taken across the X-point to estimate the inflow and outflow velocity. Running-difference images of spacetime maps from slits ``a'', ``b'', and ``c'' are used to understand the magnetic-field structuring and swirl motion on the top of the prominence. The linear slits ``a'' and ``b'' are taken along and perpendicular to the swirling motion inside prominence. The curved slit ``c'' is taken at the outer edge of the prominence to understand the magnetic-field structuring lying over the prominence.}
\end{figure*}
 A large-scale vortex motion is evident in the top part of prominence (Figure~2, movie2.mp4). The evolution of this vortex flow is observed by using SDO/AIA 304 {\AA} images at different times. The large-scale vortex flows consists of swirl motions inside (Figures~2\,--\,3, movie2.mp4). We use an array of composite images of 171+304 {\AA} to observe the spatio-temporal evolution of the swirl motion (Figure~3). Initially a linear structure is moving at the prominence-corona interface (Figure~3, middle panel). It further changes into a fragmented ring-like structure at the interface (Figure~3, bottom panel). It further merges into the southern upper part of the prominence. The typical lifetime is $\approx$ 30 minutes. The large-scale vortex motion is seen for more than two hours starting at 09:00 UT on 14 June 2018 (Figures~2\,--\,3, movie2.mp4). The condensation process of prominence has been simulated with the helical field lines in the cavity (Xia {\it et al.}, 2014). We observe that during the evolution of the prominence, it exhibits a helical structure along the spine of the prominence and plasma falls along the legs of the prominence and horns (Figures~4\,--\,5). We choose two linear slits namely ``a'' and ``b'' along and perpendicular to the swirling motion and a curved slit ``c'' along the outer edge of the prominence (Figure~6). Ten-minute running difference images at AIA 171 {\AA} are used to understand the dynamical behaviour of the plasma and possible magnetic-field structuring. During the swirling motion of the prominence, there are two types of magnetic field lines. The first type is the twisted central core, lying below the magnetic-field arcades. It consists of large-scale vortex motions and swirling motion within it. The bright features with a positive slope indicate the clockwise swirling motion on the top of the prominence (Figure~6, top and middle panels). The second type is sheared magnetic arcades, lying above the prominence in the concave shape (Figure~6, bottom panel along slit ``c''). Therefore, the present observation is consistent with the numerical model of the condensation process of prominence with the helical field lines in the cavity (Xia {\it et al.}, 2014; Fan, 2018). During the vortex motion, the sheared legs of the prominence reconnect over the neutral line, which is discussed in detail in the next sub-section. \\
\begin{figure*}
\begin{center}
\includegraphics[scale=0.8,angle=0,width=21.0cm,height=21.0cm,keepaspectratio]{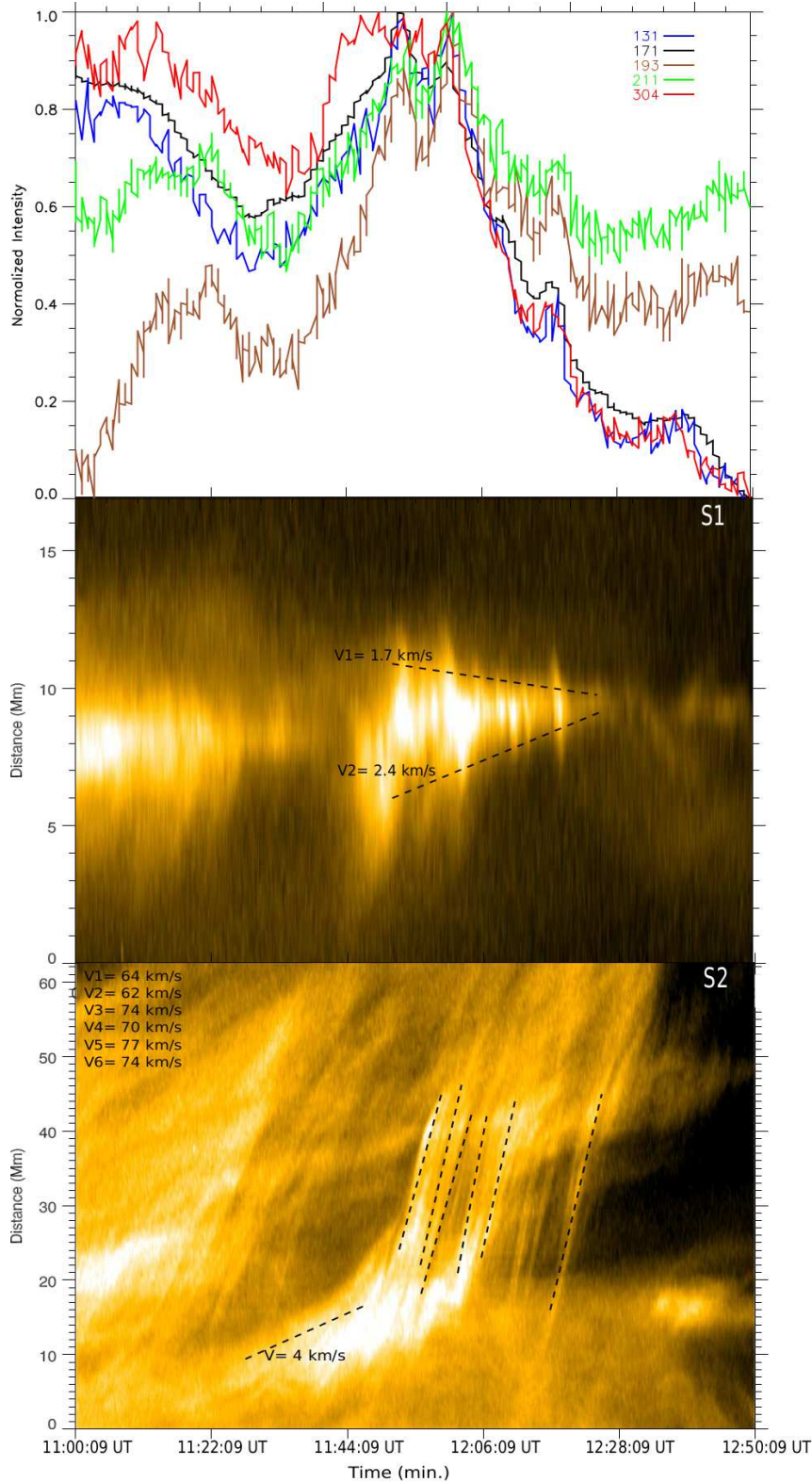}
\vspace{-0.5cm}
\caption{Top panel: The EUV emission from six different SDO/AIA filters from white box of Figure~6. The temporal variation of EUV plasma intensity (131 {\AA}, 171 {\AA}, 193 {\AA}, 211 {\AA}, 304 {\AA}, and 335 {\AA}) has estimated to understand the thermal behaviour at the reconnection site. We have analysed the inflow (rate of the thinning of the current sheet along slit S1) and outflow velocity (ejected plasmoids along slit S2) at the reconnection site. Middle panel: The thining of current sheet after reconnection gives the inflow speed at the X-point. Bottom panel: We track the path of plasmoids in H--T diagram to measure the outflow velocity.}
\end{center}
\end{figure*}

\subsection{Plasmoid Enhanced Reconnection}
Magnetic reconnection is a fundamental process that occurs in astrophysical plasmas, where magnetic field lines with anti-parallel components reconfigure. Due to magnetic reconnection the stored magnetic energy is released in the form of kinetic energy, heating, and radiation. The footpoint of the tornado, which is associated with the barbs/prominence legs, is associated with photospheric vortices. Due to shear motion at the footpoints, the legs of the prominence possess opposite magnetic-field configuration above the polarity inversion line (PIL; Figures~4\,--\,5). We observe that the southern segment of the prominence leg is rising and the northern segment is falling between 11:15 UT to 11:40 UT (Figure~5, movie3.mp4). We observe here two different kinds of magnetic reconnection. The first type of magnetic reconnection (internal reconnection) is governed by plasmoid-enhanced reconnection, which consists of an elongated current sheet (Figures~4\,--\,7). In the second stage (external reconnection), thin coronal magnetic arcades reconnect with the southern part of the overlying prominence and may be responsible for the eruption and plasma outflows. In the external magnetic reconnection an elongated thin current sheet is seen, which is mostly visible in AIA 211 {\AA}. The internal reconnection is probably triggered by the resistive tearing mode instability, which grows in the magnetically sheared legs of the prominence. The sheared field lines are destroyed above the polarity inversion line. As a result, multiple plasmoids are ejected along the height of the prominence (Figures~4\,--\,7).  \\
During the development of the prominence, oppositely directed magnetic fields associated with the prominence threads reconnect above the PIL. The southern rising segment of the prominence leg reconnects with the oppositely directed northern leg of prominence between 11:30 UT to 11:40 UT. An elongated current sheet is observed in the main body of this prominence. The ejection and acceleration of plasmoids are related to the magnetic reconnection. We observe that an elongated current-sheet is formed that lies above certain part of the polarity-inversion line of the quiet-Sun (Figure~4). Plasma downfall is also observed along the thin coronal arcades (Figure~5). The current sheet consists of multiple plasmoids (P1, P2, P3; Figure~4) ejections from the tip of the current sheet. This current sheet follows the fractional structure by the following the path: Plasmoid ejection $\Rightarrow$ thinning of current sheet $\Rightarrow$ reconnection $\Rightarrow$ plasmoid ejection $\Rightarrow$ thining of current sheet and repetition of the same process. The magnetic reconnection may create a jet, which hits and accelerate the plasmoids along the current sheet (Figures~4\,--\,7, movie3.mp4). The reconnection rate is a fundamental parameter for magnetic reconnection, which can be calculated by (e.g. Priest and Forbes, 2000; Shibata and Tanuma, 2001; Priest, 2014; Shibata and Takasao, 2016)

\begin{equation}
M=\frac{\mathrm{\textit V}_\mathrm{inflow}}{\mathrm{\textit V}_\mathrm{outflow}}
\end{equation}
here ``$M$'' is the reconnection rate, $\mathrm{\textit V}_\mathrm{inflow}$ is the inflow velocity and $\mathrm{\textit V}_\mathrm{outflow}$ is the outflow velocity. The inflow velocity may be estimated as the thinning of the current sheet (Figure~7, middle panel) and the outflow velocity is related to the plasmoid velocity (Figure~7, bottom panel). The inflow velocity is related to the apparent velocity of the reconnecting loops, which are descending so that its width is decreasing with such an apparent velocity. We estimate the inflow velocity (thinning of the current sheet along slit S1, Figure~6) to be $\approx$1.7\,--\,2.4 km s$^{-1}$ and the outflow velocity (plasmoid velocity along slit S2, Figure~6) is lying in the range of 62\,--\,77 km s$^{-1}$ (Figure~7). Correspondingly, the reconnection rate is lying between 0.02\,--\,0.05. Shibata (1995, 1996, 2001) studied the relationship between the apparent growth of flaring loop and the velocity of the plasmoids, and they found that they are correlated as (Shibata {\it et al.} 1995)
\begin{equation}
{\mathrm{\textit V}_\mathrm{plasmoid}\simeq (8 -20)\times {\mathrm{\textit V}}_\mathrm{loop}}
\end{equation}
where $\mathrm{\textit V}_\mathrm{plasmoid}$ and $\mathrm{\textit V}_\mathrm{loop}$ are the apparent velocities of the plasmoid and the flare loop, respectively. We estimate the velocity of the rising prominence and plasmoids. The prominence is rising with an apparent velocity [$\mathrm{V}$] of $\approx$ 4 km s$^{-1}$ and the apparent velocity of the plasmoids lies in the range of 62\,--\,77\, km s$^{-1}$ (Figure~7, bottom panel). We conjecture that the rising velocities of the prominence and the plasmoids are positively related as well. \\
\begin{figure*}
\hspace{-1cm}
\includegraphics[scale=0.8,angle=0,width=13.0cm,height=13.0cm,keepaspectratio]{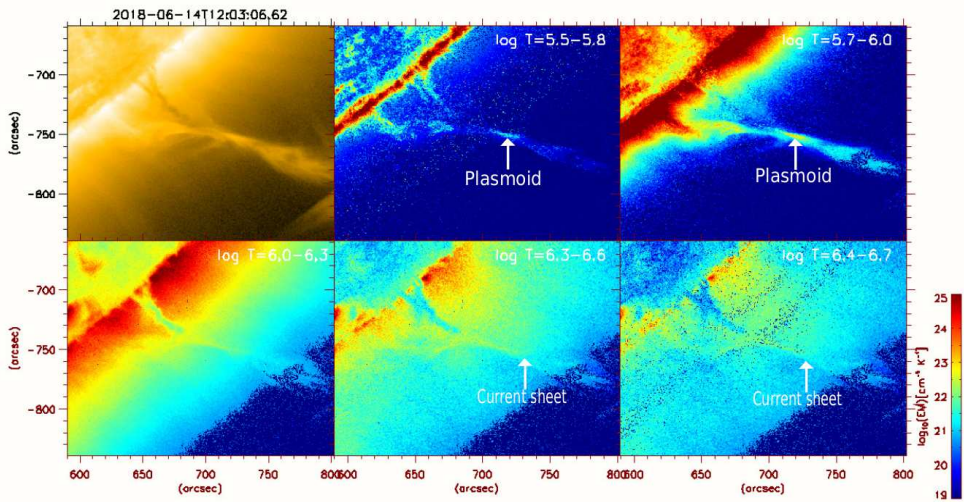}
\caption{The differential emission measure (DEM) of the sub-region to observe the thermal structure of the plasmoids and current sheet. The DEM (Cheung {\it et al.} 2015) map the temperature bins from log $T$=5.6 to log $T$=6.7 shows that plasmoid are best visible temperature range log $T$=5.8\,--\,6.1 and an elongated current sheet observed between log $T$=5.8\,--\,6.7.}
\end{figure*}
\begin{figure*}
\hspace{-1.0cm}
\includegraphics[scale=1.0,angle=0,width=13.0cm,height=13.0cm,keepaspectratio]{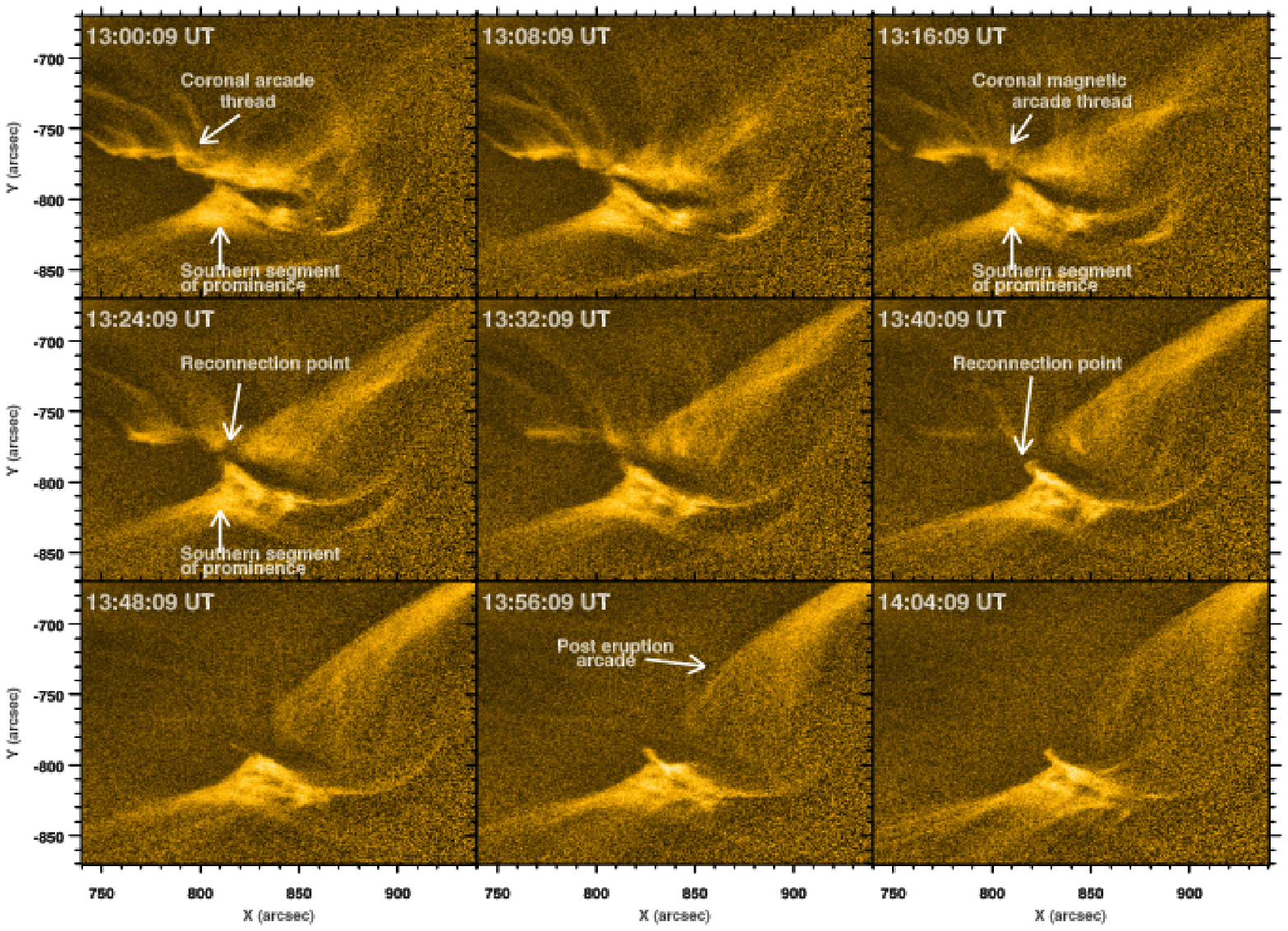}
\caption{The sub-region of the eruptive prominence after applying normalized multi-Gaussian filter to AIA 171 {\AA} imager. Several sheared magnetic arcades are evident, and they collapse near the reconnection site. A sheared coronal arcade thread reconnect with the southern segment of the overlying prominence. This external reconnection may trigger to erupt the prominence. The eruptive prominence consists of the post-eruption arcade and flux rope, which are indicated by arrows.}
\end{figure*}
We measure the temporal evolution (11:00 UT to 12:50 UT) of the emission from five different SDO/AIA filters near the reconnection X-point. The intensity at the X-point for the cool AIA 304 {\AA} (log $T$=4.7), AIA 171 {\AA} (log $T$=5.8), and AIA 131 {\AA} (log $T$=5.6) filters are higher compared to the hotter filters of AIA 193 {\AA} and AIA 211 {\AA} at 11:50 UT (Figure~7). The sheared legs of the prominence come closer and may reconnect around 11:30 UT. Oppositely directed prominence legs (the southern leg is rising and the northern segment is falling between 11:20 UT to 11:40 UT, Figure~5, movie3.mp4) reconnect above PIL. During the magnetic reconnection emissions from different filters are minimum (11:25 UT to 11:40 UT, Figure~7). The decrement in the emission is related to the coronal dimming, which may be caused by density depletion due to the lifting of eruptions in the solar corona (Sterling and Hudson, 1997; Reeves {\it et al.}, 2010; Zhu {\it et al.}; 2016). The decrement in the intensity is related to the depletion of the density. As the reconnection initiated, the prominence starts to rise and evacuate the corona. After reconnection, the first bigger plasmoid is ejected from the top of the reconnecting cusp. The intensity is maximal for the cool AIA 304 {\AA} filter and it peaks between 11:50 UT to 12:05 UT. As the reconnection begins, the other high-temperature AIA filter e.g. AIA 131 {\AA} (log $T$=5.6), AIA 171 {\AA} (log $T$= 5.8), AIA 193 {\AA} (log $T$=6.2) and AIA 211 {\AA} (log $T$=6.3), peak between 11:55 UT to 12:05 UT. The intensities are maximal near the X-point at $\approx$11:55 UT, which indicates that the spontaneous energy release occurs after the reconnection. After 12:20 UT, at some locations higher emission is seen. These higher emissions are associated with other plasmoids, which propagate along the current sheet. We use the Cheung et al. (2015) DEM method to derive the thermal properties of the prominence and the elongated current sheet. The DEM analysis indicates that plasmoids are most visible in the temperature range of log $T$(K)=5.8\,--\.6.1. A major part of the emission in this temperature range, which is lying at the million-degree corona is also observed by the SDO/AIA 171 {\AA} filter. The current sheet is best observable in the temperature range of log $T$(K)=5.8\,--\,6.7. After the reconnection local heating occurs and accelerated plasmoid ejection is observed (Figure~8). This magnetic reconnection occurs within the prominence legs, which are a partially ionized, cool, and dense plasma region. The spontaneous energy release after the reconnection may be absorbed by the surrounding plasma and may be used to accelerate the plasmoids.\\
As discussed above, plasmoid-enhanced reconnection and an elongated current sheet are observed in a prominence. The reconnection takes place between the sheared legs of the prominence above the neutral line. The resistive tearing mode most likely destroyed the sheared flux rope and multiple plasmoids are ejected (Figures~4\,--\,7). As a result multiple magnetic arcades collapse near the reconnection point. The magnetic arcades and their collapse are observed near the reconnection-point (Figure~9, top panel). The magnetic arcade may reconnect with the southern segment of the overlying prominence (Figure~9). These magnetic arcades are very thin and diffuse. Therefore, the external magnetic reconnection is not clearly observable. However, the interaction between the magnetic-arcade-associated field lines and the southern segment of the prominence is observed. After the reconnection, the magnetic arcade associated field lines move opposite to the temporary X-point and the northern segment of the prominence erupts. An elongated thin current sheet is evident above the PIL. This current sheet is associated with the external magnetic reconnection and it is best visible in the composite image of AIA 171+211 {\AA} (Figure~10). This current sheet is situated between the prominence legs and its southern segment. The reconnection rate for this external reconnection may be near the lower end of fast reconnection. Therefore, we do not observe significant inflow and heating in the current sheet. This external reconnection is responsible for the southern segment of the prominence eruption. The eruption of this prominence follows the prominence/cavity flux rope model (e.g. van Ballegooijen and Cranmer, 2010; Berger {\it et al.}, 2012). This eruptive prominence mainly consists of two parts, a core of the cool prominence plasma and hot post-eruption arcades (Figures~9\,--\,10, bottom panel). \\
\begin{figure*}
\hspace{-1.0cm}
\includegraphics[scale=1.0,angle=0,width=13.0cm,height=13.0cm,keepaspectratio]{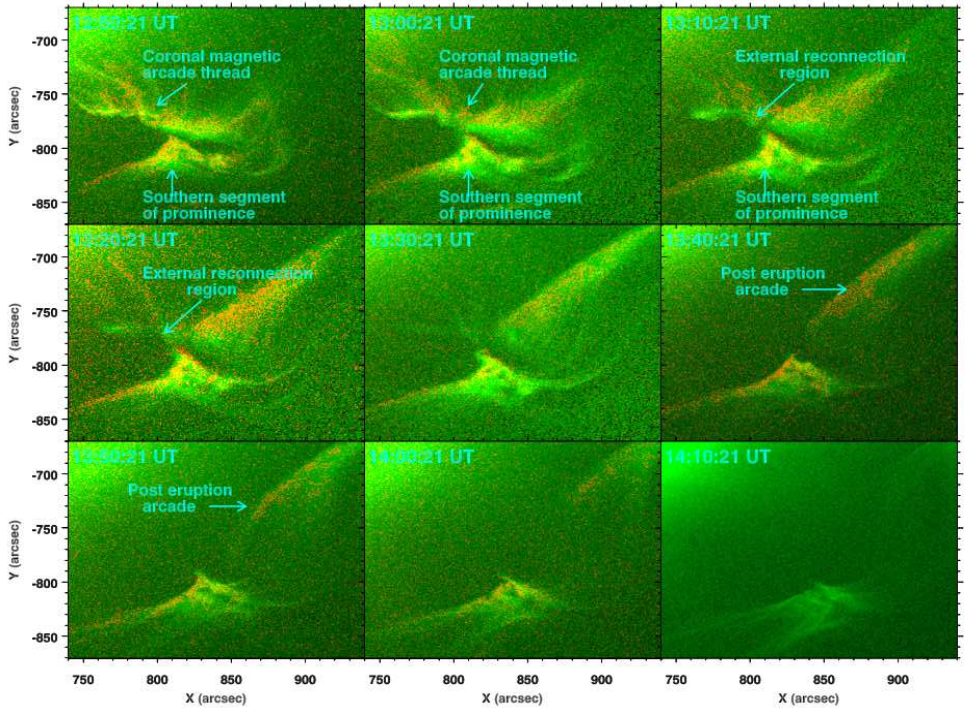}
\caption{ The sub-region of eruptive prominence as observed by the composite images of AIA 171+304 {\AA}. The composite images observed the
characteristics of hot magnetic arcades (AIA 171 {\AA}, green color) and cool prominence associated plasma (AIA 304 {\AA}, red color). A sheared coronal arcade thread reconnects with the southern segment of the overlying prominence.}
\end{figure*}

\section{Discussion and Conclusion}
A quiescent polar prominence erupted on 14 June 2018 at 13:35 UT. Before the eruption, the polar prominence showed several types of dynamics for more than five days. In the present article, we observe the development of prominence with twisting/swirling motions. The sheared prominence legs reconnect and trigger most likely the resistive tearing-mode instability. The resistive tearing-mode instability is governed by multiple plasmoid ejections within the elongated current sheet. An external magnetic reconnection takes place between the magnetic arcade and overlying prominence, which triggers the prominence eruption.\\

Several internal dynamics (barbs, spine, horns, swirl motion, and partial eruptions) are observed during the growth of this prominence and before the eruption. We investigated in details the dynamics of the prominence before and after the eruption. Cool plasma and large-scale shear) are transferred by the helical magnetic fields and horn-like structures into the prominence during its evolution (Figure~1, middle and bottom panel). We observed large-scale vortex motions in the upper part of the prominence (Figures~2\,--\,4), movie2.mp4). Twisted prominence legs transfer shear into the prominence barbs, which leads to two different kinds of magnetic reconnection above the polarity inversion line (PIL, Figures~4\,--\,9). The internal magnetic reconnection is initiated when two oppositely directed magnetically sheared prominence legs reconnect above the PIL (Figure~4, movie3.mp4). The resistive tearing-mode instability most likely is initiated and grows above the magnetic sheared layer, which destroys the sheared flux ropes. An elongated current sheet is developed above the PIL (Figures~6\,--\,8). The resistive tearing-mode instability leads to the formation of several plasmoids (Figures~4\,--\,8). These plasmoids act as an obstacle for the reconnection into the current sheet. As the plasmoids are ejected, reconnection inflow occurs. We analyzed the inflow and outflow plasma dynamics near the reconnection site (Figure~6). The velocity of the ejected plasmoids corresponds to reconnection outflow velocity and the thinning rate of the current sheet gives the inflow velocity near the X-point (Figures~6\,--\,7). We  estimated the reconnection rate, which is found to be 0.02\,--\,0.05.\\
We measured the emission of five AIA filters (131 {\AA}, 171 {\AA}, 193 {\AA}, 211 {\AA}, and 304 {\AA}) inside a box along the current sheet to observe the intensity pattern at the reconnection site (Figure~7, top panel). The intensity map indicates that cool AIA 304 {\AA} filter (log $T$=4.7) peaks at $\approx$11:50 UT. Once reconnection begins, the intensity of the higher-temperature filters, i.e. AIA 131 (log $T$=5.6, 7.0), AIA 171 (log $T$=5.8), AIA 193 (log $T$=6.2), AIA 211 (log $T$=6.3), and AIA 335 (log $T$=6.4), peak between $\approx$11:55,\--\,12:15 UT. The DEM analysis shows that the plasmoids are best visible at the coronal temperatures (log T=5.8\,--\,6.1). The elongated current sheet could be observed in the wide temperature range (log $T$=5.8\,--\,6.7; Figure~8). The emission measure and the DEM analysis indicate that the quick energy release occurred inside the current sheet in ten minutes (Figures~7\,--\,8). However, this energy release does not produce strong local heating. The reason is probably that the magnetic reconnection occurs within the prominence leg (partially ionized, dense, and cool plasma). Therefore, the energy release during the reconnection may be absorbed by the cool and dense prominence. The reconnection also triggers the overlying body of the prominence to erupt. Due to this reconnection, multiple plasmoids ejection appeared in the current sheet. The magnetic arcades start to collapse near the reconnection-point (Figure~9, upper panel). The second kind of magnetic reconnection develops between the southern segment of the prominence and collapsing magnetic arcades. A very thin, elongated, and cool (log $T$=6.3) secondary current sheet evolves, which leads to the prominence eruption (Figure~10). This current sheet is thin and cool due to its very slow reconnection rate. The eruptive prominence consists of hot post-eruption arcades and prominence associated flux rope. It disrupts the overlying coronal magnetic field. The alignment of the magnetic field and hot coronal loops are observed after this eruption (Figure~10). After the eruption, the northern segment consists of a post-eruption arcade and the southern part consists of cool prominence plasma. This eruptive prominence is associated with a CME. \\
\begin{figure*}
\hspace{-1.0cm}
\includegraphics[scale=1.0,angle=0,width=13.0cm,height=13.0cm,keepaspectratio]{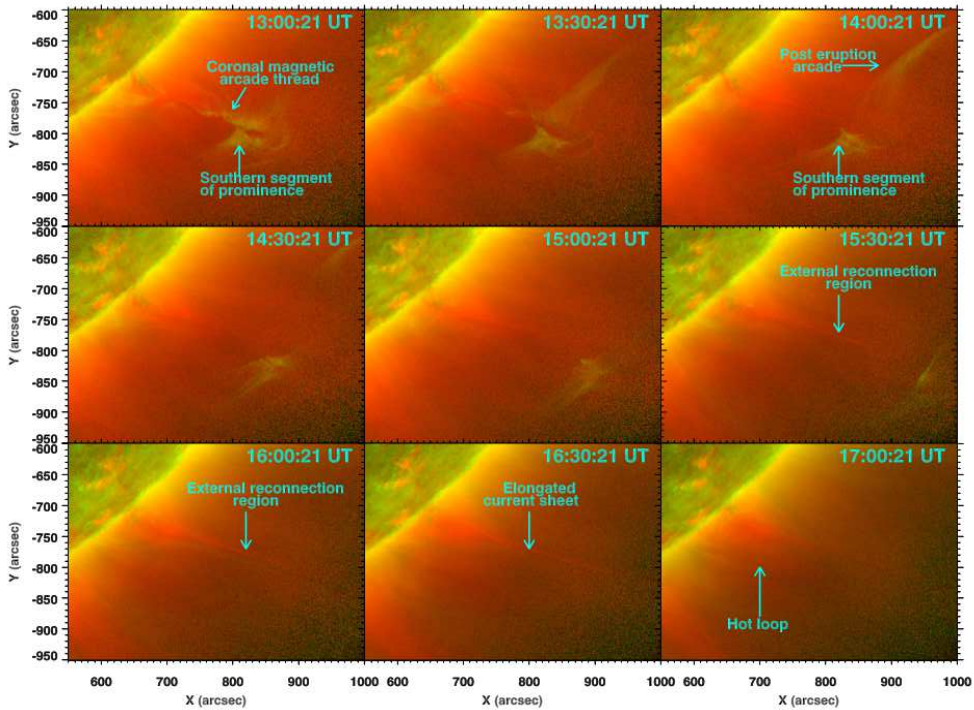}
\caption{The eruptive prominence observed as composite image of AIA 171+211 {\AA}. Southern segment of the overlying prominence reonnects with the collapsing magnetic arcades and lead to the external magnetic reconnection. A thin, elongated current sheet is evident, which is best visible in AIA 211 {\AA} (red color) waveband. The prominence-associated hot plasma, magnetic arcades and post-eruption arcades are best visible in the AIA 171 {\AA} (green color) waveband. This external magnetic reconnection triggers the prominence eruption.}
\end{figure*}
In this article, we discussed the evolution and growth of a polar prominence related to solar tornado-like motions. During the evolution of the prominence partial eruption was observed (Figure~1, movie1.mp4), but other tornado-like motions may connect to the prominence and supply cool plasma (e.g. Su {\it et al.}, 2012, 2014; Wedemeyer {\it et al.}, 2013). We observed large-scale vortex motions and swirl-like structures transferred into the overlying prominence. The swirling motion appears due to the falling plasma along the helical field lines. Pant {\it et al.} (2018) reported the tornado-like swirl motions near the footpoint of the eruptive prominence. The slow motion at the footpoint may shear the prominence legs and two-stage reconnection takes place (e.g. Kusano {\it et al.}, 2003, 2004; Nemati {\it et al.}, 2015). The similar dynamics observed in this prominence, in which sheared and oppositely directed prominence legs reconnect over a PIL (Figures~4\,--\,6, movie3.mp4) and multiple plasmoids ejection are observed (Figures~4\,--\,6). The estimated magnetic reconnection rate initiated by plasmoid enhanced reconnection is consistent with the Petscheck-type reconnection (e.g. Priest and Forbes, 2000; Shibata {\it et al.}, 2001). The external reconnection may be responsible for prominence eruption (Figure~7). The eruptive quiescent prominence consists of hot post-eruption arcades and a flux rope associated cool plasma structure (Figures~8\,--\,9), which is consistent with the van Ballegooijen and Cranmer (2010) and Berger (2012) coronal cavity/prominence flux rope model.\\

The present work to the best of our knowledge provides a relation between tornado-like motions, formation, and eruption of prominence through plasmoid enhanced reconnection. The tornadoes transport the cool plasma and sheared magnetic fields in the prominence, and they may be responsible for the generation of tearing mode instability, thining of current sheet and ejection of plasmoids. The prominence system is also subjected to the two stage magnetic reconnection during these whole physical processes. Later, the external magnetic reconnection in the upper part of the prominence system leads to its eruption and the associated plasma outflows. These physical processes develop at different spatio-temporal scales in a sequential manner. Therefore, overall they provide the complete scenario of the formation of polar prominences, the associated large-scale vortex motions, and the formation of elongated current-sheet through a series of reconnection events.

\section*{Acknowledgments}
The authors thank the anonymous reviewer for valuable comments that have helped us to substantially improve the manuscript. A.K. Srivastava and S.K. Mishra acknowledge the DST-SERB (YSS/2015/000621) project. A.K. Srivastava acknowledges the UKIERI Research Grant for the support of his research. P.F. Chen was supported by the Chinese grants NSFC 11533005, 11961131002 and Jiangsu 333  Project (No. BRA2017359). We acknowledge the use of Cheung {\it et al.} (2015) for calculating the differential emission measure (DEM). Data courtesy of NASA/SDO and the AIA science team.

\section*{Disclosure of Potential Conflicts of Interest}
The authors declare that they have no conflicts of interest.

\end{document}